\documentclass{aa}  
\authorrunning{Zielinski et al.}
\usepackage{graphicx}
\usepackage{color}
\usepackage{upgreek}
\usepackage{footnote}
\makesavenoteenv{tabular}
\usepackage{txfonts}

\begin{document}

   \title{Constraining the magnetic field properties of Bok globule B335 using SOFIA/HAWC+}
   \titlerunning{Magnetic field properties of the Bok globule B335}


   \author{
   N. Zielinski, S. Wolf and R. Brunngräber
          }
    \authorrunning{N. Zielinski}

   \institute{University of Kiel, Institute of Theoretical Physics and Astrophysics, Leibnizstrasse 15, 24118 Kiel, Germany\\
              \email{nzielinski@astrophysik.uni-kiel.de}}

   \date{Received / Accepted }

 
  \abstract
 {Thanks to their well-defined shape and mostly isolated locations, Bok globules are suitable objects for studying the physics of low-mass star formation. To study the magnetic field of the prototypical Bok globule B335, we obtained a spatially resolved polarization map with SOFIA/HAWC+ at a wavelength of 214\,$\mu$m. For the first time, these observations reveal that polarization holes in Bok globules, that is, the decrease in polarization degree towards their dense centers, also occur in the far-infrared wavelength regime. The observed polarization pattern is uniform with a mean polarization angle of 48$^\circ\pm$\,26$^\circ$ and a magnetic field strength of $\sim$ 142$\,\mu$G.  Moreover, we use complementary polarimetic data for B335 obtained  at near-infrared to millimeter wavelengths to analyze and constrain the magnetic field across different scales. By applying the 3D Monte-Carlo radiative transfer code POLARIS \citep{Reissl2016}, we developed a model for the density and magnetic field structure as well as for the dust properties of this globule. We conclude that the column density towards the center of B335 is too low to cause the observed polarization hole in B335 via dichroic absorption \citep{BrauerWolfReissl}. Furthermore, we conclude that the effect of self-scattering has no significant impact on the observed polarization. Adopting dust-grain alignment via the radiative torque mechanism, a combination of the interstellar radiation field and the central star as radiation sources is consistent with the decrease in polarization degree at the outer regions of B335 ($\approx$\,10$^4$\,au from the core). However, the model fails to explain the low polarization degree within the inner 5\,000\,au.}


   \maketitle
%

\section{Introduction}
Stars form in dense accumulations of interstellar dust and gas. However, our understanding of the star formation process, including the roles that various physical processes play during the evolution of a collapsing molecular cloud to an evolved star, remains incomplete. In particular, the impact that  magnetic fields may have at various stages of the star formation process is a matter of open debate \citep{Matthews2002, Pudritz2013, Seifried2015}. For example, magnetic fields are considered to slow down the contraction of star-forming regions and filaments,  thus providing  a possible explanation of the observed low star formation rates \citep{VanLoo2015, Federrath2015}. Furthermore, such fields are expected to influence the coupling of the gas and dust phase as well as  the shape of cloud fragments \citep{Wolf2003}. Measuring magnetic field strengths and structures is the most direct approach to addressing these open issues. One way to obtain this information is through polarimetric observations of the thermal re-emission radiation.

It is widely assumed that the polarization observed in Bok Globules at submillimeter/millimeter (submm/mm) wavelengths is caused by aligned elongated dust grains. This provides the opportunity to indirectly observe the magnetic field if the dust grains are aligned with the magnetic field direction. Several alignment theories exist \citep[for an overview see, e.g.,][]{Andersson2015}. One of the most promising is based on radiative torques (RATs). Here, elongated dust grains are assumed to spin up and align with their longer axis perpendicular to  the magnetic field lines because of the Barnett effect \citep{Barnett1915, Lazarian2007, LazarianHoang2007, Hoang2009} in the presence of an anisotropic radiation field (e.g., central star embedded in a circumstellar envelope). However, there is a competing process that also causes polarized emission but is not associated with the magnetic field, that of self-scattering \citep{Kataoka2015}. Here, the re-emitted thermal radiation of the dust grains gets scattered by the dust and thus polarized as well.

Assuming the detected polarization is caused by the emission of magnetically aligned nonspherical dust grains, it is possible to draw conclusions about the underlying magnetic field.
Analyzing the polarization pattern provides information about the magnetic field strength and structure. Additionally, a strong magnetic field might lead to a close alignment between outflow and magnetic field direction \citep{Matsumoto2006}. 
Furthermore, with knowledge of the magnetic field strength, one can evaluate whether the magnetic field is strong enough to provide support against gravity due to magnetic pressure \citep{Crutcher2004}.  \\

Bok globules are suitable objects to study star formation because they are compact, mostly isolated objects with simple shapes and show signs of 
 low-mass star formation \citep{Bok1977}. Furthermore, polarimetric observations of these objects  reveal a peculiar property of their polarization pattern at submillimeter wavelengths: In various cases, Bok globules show a decrease in polarization degree towards their  dense central regions \citep[``polarization holes'', e.g.,][]{Henning2001, Wolf2003}. 
Several hypotheses exist that try to explain this phenomenon: \textbf{(a)} an increased disalignment of the dust grains due to higher density and temperature \citep{Goodman1992},  \textbf{(b)} an insufficient angular resolution of a possibly complex magnetic field structure on scales below the resolution of the polarization maps \citep[e.g.,][]{Shu1987,Wolf2004},  \textbf{(c)} less elongated dust grains in dense regions \citep[e.g.,][]{Creese1995, Goodman1995}, and (\textbf{d)} unaligned graphite grains accumulated in dense regions \citep[e.g.,][]{Hildebrand1999}.

In addition to the effects quoted above, \citet{BrauerWolfReissl} concluded that \textbf{(e)} certain combinations of optical depth,  dust grain size, and chemical composition may also provide an explanation for the phenomenon of polarization holes.  Here, two scenarios are of interest: As long as the globule is optically thin at the observing wavelength, the net polarization of the polarized thermal re-emission radiation of aligned nonspherical dust grains increases with increasing column density. However, once absorption becomes relevant, for example, towards the central regions of dense molecular cloud cores,  dichroic extinction will decrease the degree of polarization towards the globule core. This effect may even cause a flip of the polarization direction by 90$^\circ$ if the dichroic extinction dominates over dichroic emission \citep{Reissl2014, BrauerWolfReissl}, as seen in NGC1333 IRAS4A \citep{Ko2020}.

In order to test the hypothesis of the interaction between dichroic extinction and emission, observations at several wavelengths are required to constrain the properties of the dust. This allows us to constrain the wavelength-dependent optical depth, and thus to assess the relative contribution of both counteracting polarization mechanisms, that is,  dichroic extinction versus polarized emission.
The Bok globule B335 is a particularly suitable object for this study because high-quality data are available from existing polarimetric observations ranging from the near-IR \citep{Bertrang2014} to submm/mm wavelengths \citep[e.g.,][]{Wolf2003, Maury2018, Yen2019, Yen2020}. \\

This paper is organized as follows: In Section \ref{Section_Observation}, we describe the data acquisition and reduction, and the selection criteria we used to constrain the data. In Section \ref{Section_Results} we present the polarization map of B335, our analysis, and the corresponding model. We discuss the origin of the observed polarization and present the derived magnetic field structure and strength of B335 (Sect. \ref{Section:Self_Scattering} and following). 
Additionally, in Section \ref{Section:HAWC_plus_observations_of_B335_in_the_context_of_further_polarimetric_observations} we provide a short discussion of our findings for the magnetic field structure  in the context of other polarimetric observations. Furthermore, we discuss possible reasons for the observed polarization hole in B335 (Sect. \ref{Section:Polarization_hole_in_B335}). Finally, we summarize our results in Section \ref{Section_Conclusion}.

\section{Observations}  \label{Section_Observation}

\subsection{Source description}

B335, a  Bok  globule  at  a  distance  of  $\approx$ 100 pc \citep{Olofsson2009}, is one of the best studied molecular cloud cores. An east--west elongated molecular outflow with a dynamical age \mbox{of $\approx 3 \cdot 10^4$ yr} \citep[e.g.,][]{Chandler1993}  was observed by \citet{Hirano1988}. The core of B335 is slowly rotating \citep{Kurono2013}. Its size was determined to be $\approx$ 0.1\,pc \citep{Motte2001, Shirley2002}. The mass of the protostar was estimated to be $\approx 0.4\,$M$_\odot$ \citep{Zhou1993, Choi1995}. \citet{Wolf2003} derived a total envelope mass of $\approx$ 5$\,$M$_\odot$ within a radius of 1.5 $\cdot$ 10$^4$ au from JCMT/SCUBA observations. With single-dish and interferometric observations, infall motions have been detected \citep[e.g.,][]{Zhou1993, Kurono2013}. The simple structure of the object \citep[e.g., ][]{Wolf2003, Davidson2011} makes it an excellent candidate to study the implications of magnetic fields in collapsing dense cores.

\subsection{Data acquisition} \label{Section:Data_Aquisition}

SOFIA/HAWC+ band E observations of the globule B335 were carried out on 18 October 2017 as part of the SOFIA Cycle 5 (Proposal 05$\_$0189). Band E provides an angular resolution of 18.2$''$ full width at half maximum (FWHM) at the 214 $\mu$m center wavelength. The detector format consists of two 64 x 40 arrays, each comprising two 32 x 40 sub-arrays \citep{Harper2019}. The observations were performed using the chop-nod procedure with a chopping frequency of 10.2\, Hz.

The raw data were processed by the HAWC+ instrument team using the data reduction pipeline version 1.3.0. This pipeline consists of different data processing steps including corrections for dead pixels as well as the  intrinsic polarization of the instrument and telescope  \citep[for a brief description of all steps, see e.g.,][]{Santos2019} resulting in “Level 4” (science-quality) data. These include FITS images of the total intensity (Stokes $I$),   polarization degree $P$, polarization angle $\theta$, Stokes $Q$ and Stokes $U$, and all related uncertainties. Furthermore, to increase the reliability of our findings, we apply two additional criteria for the data that will be considered in the subsequent analysis:
\begin{equation}  \label{Formula:Requirement1}
 \hspace{3cm} \frac{I}{\sigma_I}  > 100 \: ,
\end{equation}
\begin{equation} \label{Formula:Requirement2}
\hspace{3cm}  \frac{P}{\sigma_P}  > 2.5 \: ,
\end{equation}
where $\sigma_I$ and $\sigma_P$ are the standard deviations of $I$ and $P$, respectively. The 19 data points listed in Table \ref{Table:SOFIA_Detections} fulfill these requirements.


\section{Results} \label{Section_Results}

\subsection{Polarization map of B335 (at 214$\,\mu$m) }
\begin{figure}
   \centering
   \includegraphics[width=\hsize]{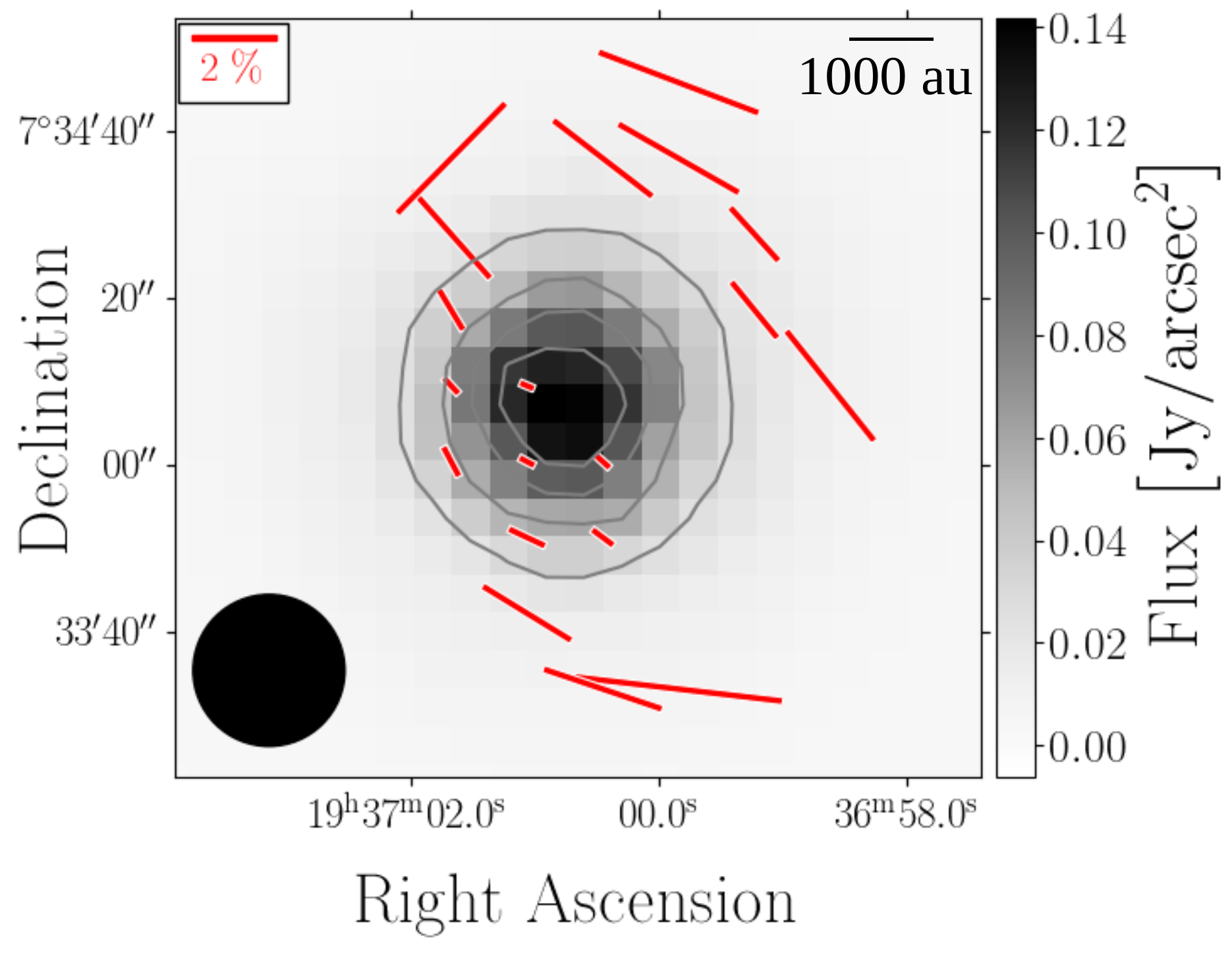}
      \caption{SOFIA/HAWC+ band E (214\,$\mu$m) polarization map of B335. Total intensity is shown with overlaid polarization vectors in red. The length of the vectors is proportional to the polarization degree and the direction gives the orientation of the linear polarization. The isocontour lines mark 20\%, 40\%, 60\%,\ and 80$\%$ of the maximum intensity. According to criteria (\ref{Formula:Requirement1}) and (\ref{Formula:Requirement2}) only vectors with $I > 100\, \sigma_I$ and \mbox{$P > 2.5\, \sigma_P$} are considered (see Sect. \ref{Section:Data_Aquisition}). The beam size of 18.2$''$ (defined by the FWHM)  is indicated in the lower left.
              }
         \label{Pol_Map_Only_Valid_Vectors}
\end{figure}

Figure \ref{Pol_Map_Only_Valid_Vectors} shows the resulting 214$\,\mu$m polarization map of B335 overlaid on the corresponding intensity map. The peak emission amounts to 0.14\,Jy/arcsec$^2$. In those regions where criteria (\ref{Formula:Requirement1}) and (\ref{Formula:Requirement2}) concerning the signal-to-noise ratio (S/N) are fulfilled (see Sect. \ref{Section:Data_Aquisition}), the  polarization degree decreases from $\sim$5\,$\%$ at the outer regions to $\sim$0.5\,$\%$ at the center. The decrease of the polarization degree towards the center has also been observed in earlier studies in the case of various other Bok globules \citep[e.g.,][]{Wolf2003}.  Despite the increasing flux towards the central region, the data in the north-eastern area of the central region of the globule do not fulfill criterion (\ref{Formula:Requirement2}). However, at the same time this finding provides an upper limit for the polarization degree of 0.25$\,\%$ in this particular region.  Thus, this observation demonstrates for the first time that polarization holes in Bok globules may also occur in the far-infrared (far-IR) wavelength regime. The orientation of the polarization vectors shows a highly uniform pattern with a mean polarization \mbox{angle of 48$^\circ$.} \\

\subsection{Distribution of the polarization vectors and the polarization degrees} \label{Section:Distribution_pol_vectors_pol_degree}
\begin{figure}
\includegraphics[width=\hsize]{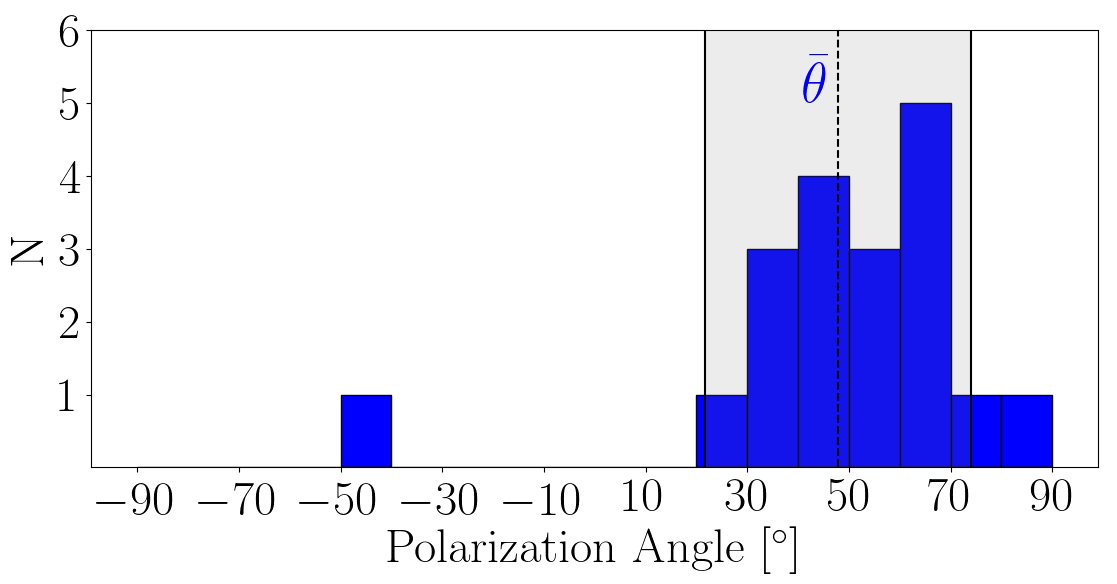}
 \includegraphics[width=\hsize]{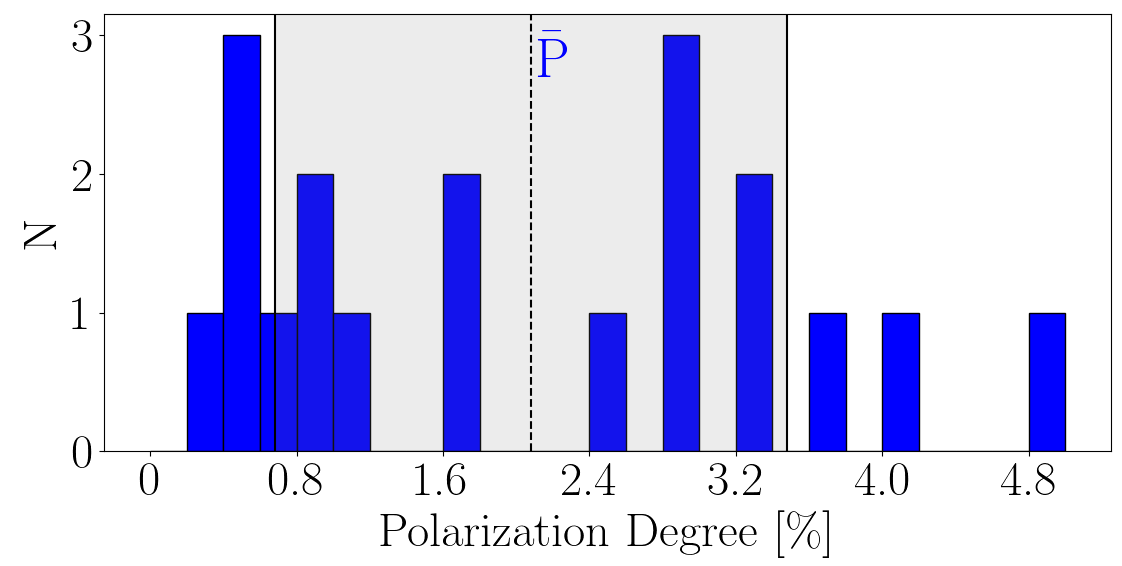}
      \caption{Histogram showing the distribution of polarization angles (top) and polarization degrees (bottom) in the far-IR obtained with SOFIA/HAWC+. The dashed lines represent the mean polarization angle $\bar{\theta} = 48^\circ$ and mean polarization degree $\bar{P}$ = 2.1$\,\%$, respectively. The solid lines represent the corresponding 1$\sigma$ levels, 26$^\circ$ , and 1.4\,$\%$, respectively. 
              }
              \label{Figure:Distribution_Pol_Angles_Pol_Degrees}
\end{figure}

Figure \ref{Figure:Distribution_Pol_Angles_Pol_Degrees} (top) shows the distribution of the polarization angle $\theta$. We mostly find polarization angles ranging from 20$^\circ$ to 80$^\circ$ with clear predominance around 60$^\circ$. 

\begin{figure}[h]
   \centering
   \includegraphics[width=\hsize]{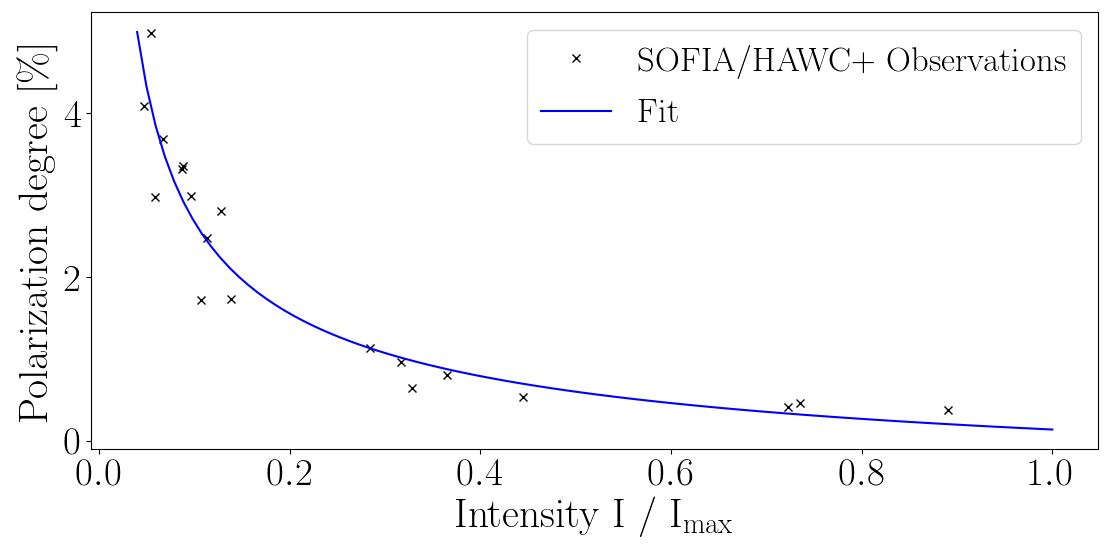}
      \caption{Correlation between the polarization degree and the normalized intensity $I$ (see Sect. \ref{Section:Distribution_pol_vectors_pol_degree} for details). 
              }
         \label{Scatter_Plot_B335}
\end{figure}

 The distribution of the polarization degree $P$ is shown in \mbox{Fig. \ref{Figure:Distribution_Pol_Angles_Pol_Degrees} (bottom). }
The degree of polarization varies between 5$\,\%$ and $\sim$\,0.5$\,\%$. In the following, we analyze $P(I)$, with intensity normalized to the maximum intensity $I_\text{max}$ (see Fig. \ref{Scatter_Plot_B335}).  Assuming optically thin emission, the relationship between polarization degree $P$ and intensity $I$ is equivalent to the relation between $P$ and the corresponding column density.

Inspired by the work of \citet{Davis2000}, who applied linear least-square fits to polarimetric observations of the Serpens cloud core and found a correlation between the measured polarization and intensity, \citet{Henning2001} found a correlation between these quantities as well in the case of the two Bok globules CB54 and DC253-1.6. 

They approximated the decrease of the polarization degree as a function of increasing intensity by
\begin{equation}
\hspace{3cm}  P = a_0 + a_1 \cdot \left( \frac{I}{I_\text{max}} \right)^{a_2}, 
\end{equation}
where  $a_0 , a_1$, and $a_2$ are fitting parameters. For the 214\,$\mu$m polarization map of B335, we derive following values for these parameters: $a_0 \, {=} \, -0.85, a_1 \, {=} \, 0.99$, and $a_2 \, {=} \, -0.55$. A measure of the goodness of the fit is given by the coefficient of determination $R^2$ = 90.5$\,\%$. 
\citet{Wolf2003} performed a similar analysis for the polarization holes of the globules B335, CB230, and CB244 at a wavelength of 850$\,\mu$m. As a result, they obtained  averaged values of the fit parameters \mbox{$a_0 \, {=} \, -1.7,a_1 \, {=} \, 3.96$,} and $a_2 \, {=} \, -0.43$. Thus, the parameter  $a_2$ which describes the slope of the polarization degree is similar at both wavelengths.

\subsection{Model of B335} \label{Section:Model_of_B335}
To provide a basis for a profound analysis of the polarization maps, we first derive constraints on the structure, mass, and dust properties of the Bok globule B335. For this purpose, we use a model based on a Bonnor-Ebert density distribution, which is motivated by the rather simple circular shape of the object. For the same reason, it is often used in modeling studies of Bok globules \citep[e.g.,][]{Harvey2001, Kandori2005, BrauerWolfReissl, Kandori2020}. The radially dependent density structure is given by
\begin{equation}
\hspace{2cm}  \rho(r) = \begin{cases}

\quad \rho_0 \cdot R_0^{-2}, & \text{if} \quad r \le R_0  \\ 

\quad \rho_0 \cdot r^{-2}, & \text{if} \quad R_0 < r \le R_\text{out} \\

\quad 0, & \text{if} \quad r > R_\text{out}.

\end{cases}
\end{equation}
 
The density structure is characterized by the truncation radius $R_0$, defining an inner region with constant density. The truncation radius is a free parameter and we obtain its value below by fitting the radial brightness profile of B335. 
Outside this region, the density decreases towards the outer radius $R_\text{out}$. Following \citet{Kandori2020}, we use a value of \mbox{ $R_\text{out}$ = 13\,100 au}. 
For the central heating source we apply a pre-main sequence star \mbox{(T Tauri star)} with a luminosity of $L_\star$ = 1.5\,L$_\odot$ and a radius of $R_\star$ = 1.5\,R$_\odot$. These values were introduced by  \citet{Evans2015} to calculate the mass infall rate of B335. \\
We adopt compact, homogeneous, spherical dust grains consisting of silicate and graphite \citep{Olofsson2011} with a mass abundance ratio of 62.5$\%$ silicate and 37.5$\%$ graphite\footnote{Due to the anisotropy of graphite, the 1/3 - 2/3 approximation \citep{Draine1984,Draine1993} is applied.}. We apply the canonical gas-to-dust ratio of 100:1 and assume that the grain sizes are distributed similarly to the power-law distribution of the ISM:
\begin{equation}
\hspace{2cm}  \text{d}n(a) \propto a^{-3.5} \: \text{d}a \: , \qquad a_\text{min} < a < a_\text{max},
\end{equation}
where d$n(a)$ represents the number of grains with a radius in the range $[a, a+\text{d}a]$ \citep{MathisRumplNordsieck1977}. While we set the minimum dust grain size to $a_\text{min}$ = 5\,nm, as often used for the description of the dust phase of the ISM, it is questionable whether the same argument applies for the maximum grain size  \citep[$a_\text{max}$ = 250\,nm for the ISM; see e.g.,][]{MathisRumplNordsieck1977}.
On the one hand, Bok globules represent an early stage of star formation with densities that -- in contrast to the conditions in protoplanetary disks -- are not favorable for grain growth. On the other hand, \citet{Yen2020} proposed the presence of grains with sizes of a few tens of $\mu$m  in B335 based on 0.87\,mm ALMA observations, but on a size scale of $\sim$ 100 au, that is, on scales significantly smaller than representive for our SOFIA observation. However, given this uncertainty,  we consider maximum dust grain sizes for our model in the range of 250\,nm - 50$\,\mu$m. An overview of the considered parameters and chosen values for our model is given in Table \ref{Table_model_parameter_b335}. \\

\begin{table}
  \begin{center}
    \caption{Overview of the parameters considered in the Bok globule model for B335. Free parameters are marked with the symbol  "$\updownarrow$".}
    \label{Table_model_parameter_b335}
    \begin{tabular}{lcr}
    \hline \hline    \rule{0pt}{2ex}
      \textbf{Parameter} & \textbf{Symbol} & \textbf{Value}\\
      \hline \hline
      \rule{0pt}{3ex}
      \noindent \textbf{Globule structure} \\
       \rule{0pt}{1ex}
        Outer radius  & $R_\text{out}$ $^a$ & 13\,100$\,$au\\
      \rule{0pt}{1ex}
      Truncation radius  & $R_0$ ($\updownarrow$) & 1\,000$\,$au\\
       \rule{0pt}{1ex}
      Gas mass & $M_\text{gas}$ ($\updownarrow$) & 1-5$\,$M$_\odot$\\
       \rule{0pt}{3ex}
      \textbf{Central star} \\
       \rule{0pt}{1ex}
      Luminosity  & $L_\star$ $^b$ & 1.5$\,$L$_\odot$\\
       \rule{0pt}{1ex}
      Radius  & $R_\star$ $^b$ & 1.5$\,$R$_\odot$\\
       \rule{0pt}{3ex}
      \textbf{Dust properties}\\
       \rule{0pt}{1ex}
      Minimum grain radius & $a_\text{min}$ & 5$\,$nm\\
       \rule{0pt}{1ex}
      Maximum grain radius & $a_\text{max}$ ($\updownarrow$) & 250\,nm - 50$\,\mu$m\\
      \hline \hline
    \end{tabular}
     \tablefoot{\footnotesize $^a$\citealt{Kandori2020}; $^b$\citealt{Evans2015}}
  \end{center}
\end{table}

In the following, we constrain the mass of B335 and the maximum grain size by fitting the observed radial brightness profile at $\lambda$ = 214\,$\mu$m. For this purpose, we perform radiative transfer simulations  using the publicly available Monte Carlo 3D radiative transfer code POLARIS\footnote{http://www1.astrophysik.uni-kiel.de/$\sim$polaris/} \citep{Reissl2016}. We find that a truncation radius of $R_0$ = 1000\,au provides the best agreement between the shape of the derived radial brightness profile and the observed one. First, the temperature profile of B335 is calculated self-consistently on the basis of the embedded radiation source and the dust distribution. Subsequently,  the re-emission maps are computed based the temperature maps. The procedure is as follows: For a given maximum dust grain size, the globule mass is adjusted to match the observed peak emission. 
Therefore, in this case, the globule mass is closely connected to the maximum dust grain size. This yields different combinations of maximum dust grain sizes and globule masses.  Constrained by the range of possible maximum grain radii $a_\text{max} \in$ [250\,nm, 50\,$\mu$m], the corresponding globule mass range is $M_\text{gas} \in$ [1, 5] M$_\odot$.

\begin{figure}
   \centering \includegraphics[width=\hsize]{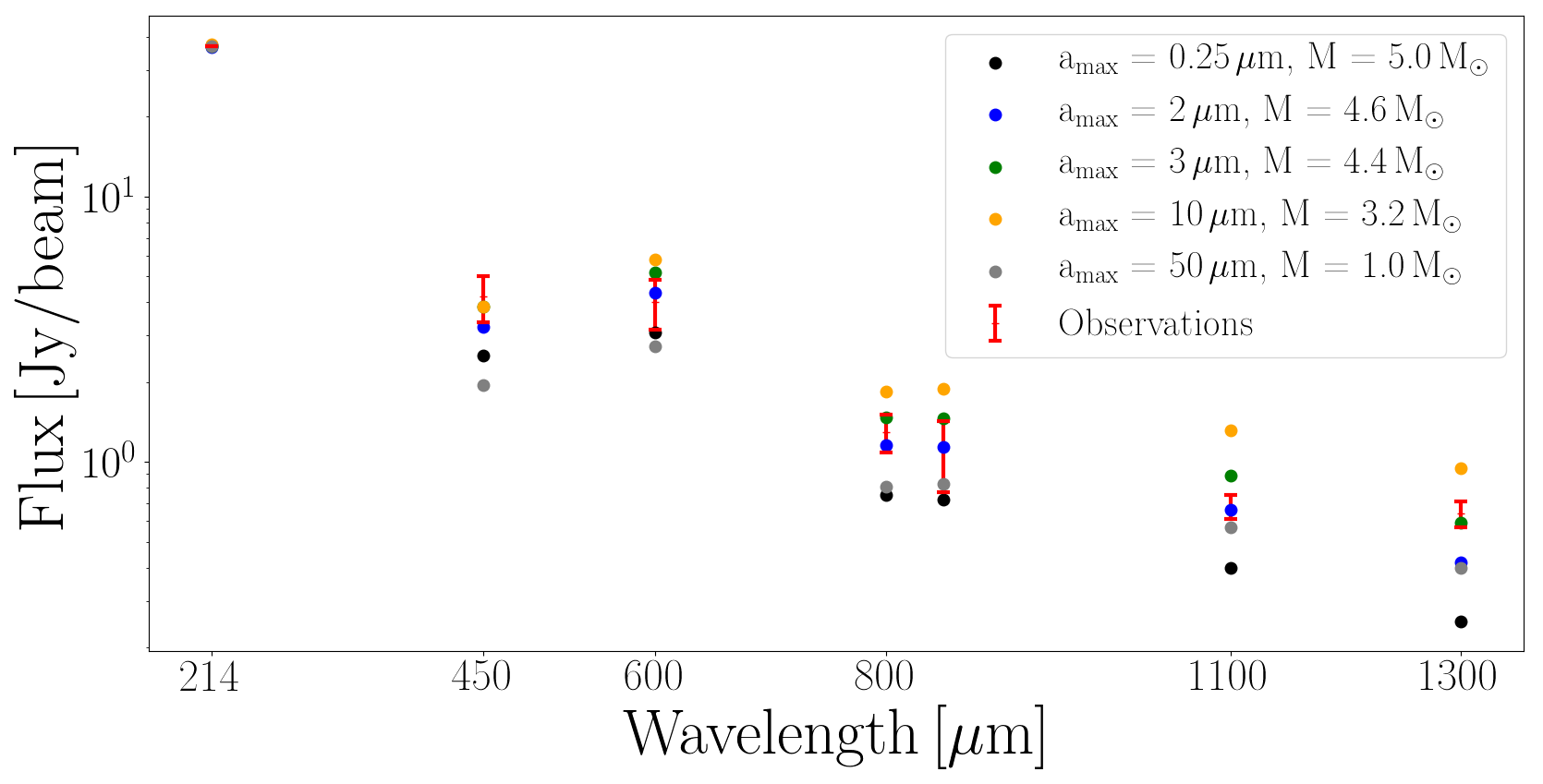} \caption{Overview of observed fluxes and the corresponding fluxes resulting from POLARIS simulations for combinations of maximum grain sizes $a_\text{max} \in$ [0.25, 50]\,$\mu$m and globule masses $M_\text{gas} \in$ [1.0, 5.0]\,M$_\odot$.} \label{Overview_observed_Fluxes_Simulations}
      \centering
   \includegraphics[width=\hsize]{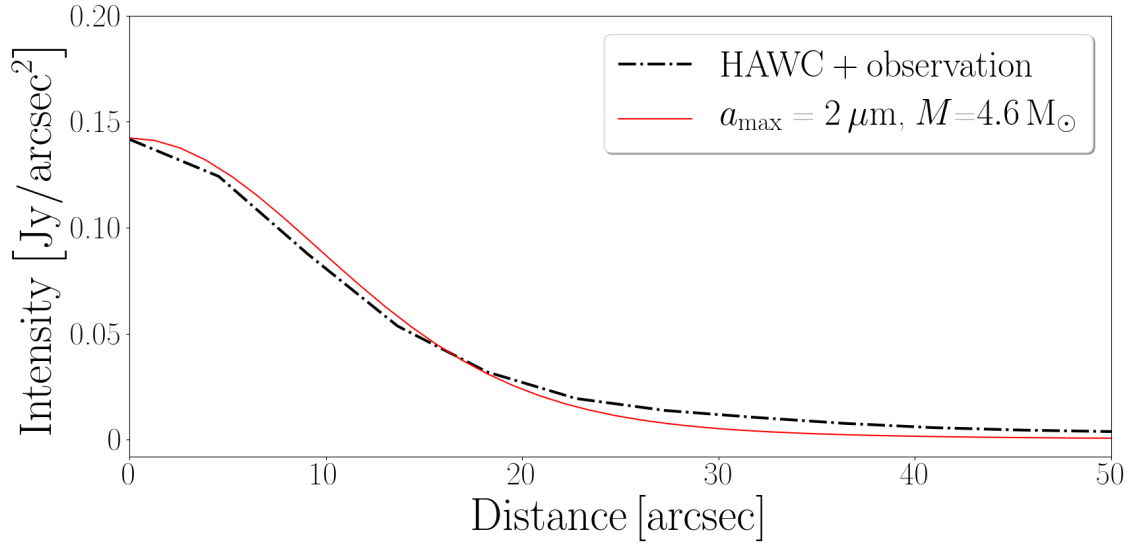}
      \caption{ Radial profile ($I$ vs. $r$) for $a_\text{max}$ = 2\,$\mu$m and $M_\text{gas}$ = 4.6\,M$_\odot$ in comparison to the observed profile at $\lambda$ = 214$\,\mu$m.
              }
         \label{Radial_Profile_I_vs_r_250nm_to_50_micron}
\end{figure}

To further constrain the parameters $a_\text{max}$ and $M_\text{gas}$, we compare the numerically derived fluxes with observational results at a wide range of wavelengths obtained with single-dish telescopes with a resolution similar to that of SOFIA/HAWC+. In particular, B335 was observed with the 15-m James Clerk Maxwell Telescope at 450$\,\mu$m, 600$\,\mu$m, 800$\,\mu$m, 850$\,\mu$m, 1100$\,\mu$m, and 1300$\,\mu$m \citep{Chandler1990, Launhardt2010}.
 The result is given in Fig. \ref{Overview_observed_Fluxes_Simulations} and Table \ref{Table:Overview_Comparison_Observation_Simulation}.  The model with a maximum dust grain size of $a_\text{max}$ = 2$\,\mu$m and a corresponding mass of $M_\text{gas}$ = 4.6\,M$_\odot$  best fits the observations  (\mbox{$\chi^2$ = 0.35}, see Table \ref{Table:Overview_Comparison_Observation_Simulation} and  Fig. \ref{Radial_Profile_I_vs_r_250nm_to_50_micron} for the corresponding radial brightness profile). The resulting best-fit model has an optical depth of only $\tau_{\text{214}\mu\text{m}} \approx 0.1$ towards the center. \\

An alternative and commonly used approach to derive the mass of interstellar clouds relies on the assumptions of optically thin dust emission and constant temperature. Here, one calculates the hydrogen gas mass $M_H$ based on the measured flux density by
\begin{equation}
M_\text{H} = M_\text{dust} \: X = \frac{S_{\nu} \, D^2 }{\kappa_d(\nu) \, B_\nu (\nu, T_d)} \: X \: ,
\end{equation}
 where X is the hydrogen-to-dust mass ratio ($X$ = 100), $ M_\text{dust} $ the dust mass, $S_\nu$ the observed total flux density, $D$ the distance to the object, $\kappa_\text{d}(\nu)$ the dust opacity, $B_\nu (\nu, T_\text{d})$ the Planck-function, $\nu$ the frequency, and $T_\text{d}$ the dust temperature. Making use of the best-fit model derived above, we obtain a mean dust temperature $T_\text{d} \sim$ 14\,K  by averaging the resulting temperature of our temperature map-simulation inside a circle around the peak value with a radius of 3$\sigma$\footnote{The value for $\sigma$ (standard deviation) is obtained by fitting a Gaussian to the radial brightness profile of B335.}. The derived temperature is similar to literature values \citep[10 - 14\,K; see][]{Wolf2003}.
 For the considered dust model (with $a_\text{max}$ = 2\,$\mu$m), the dust opacity amounts to $\kappa_\text{d}(\,214\,\mu\text{m}) \approx$ 0.99 m$^2$ kg$^{-1}$. This results in a hydrogen mass of $\approx$ 3.5\,M$_\odot$. Considering helium and heavy elements, the total gas mass amounts to $M_\text{gas}$ $\approx$ 1.36 $M_\text{H}$ $\approx$ 4.7\,M$_\odot$,  which is very close to the value derived without the additional simplifications. 
We  conclude that the derived mass based on the 214\,$\mu$m SOFIA/HAWC+ observation is comparable to values published in earlier studies \citep[$\approx$ 3.5 - 5\,M$_\odot$;][]{Wolf2003, Kandori2020}. \\

\subsection{Origin of the polarization at 214$\,\mu$m: Polarization due to self-scattering in the context of Bok globules?} \label{Section:Self_Scattering}

Until a few years ago, it was generally accepted that the (sub)millimeter polarization observed in Bok globules was caused by the emission or absorption of magnetically aligned nonspherical dust grains. However, as spatially resolved polarization maps of selected circumstellar disks \citep[e.g., HL Tau, ][]{Stephens2017} in the same wavelength range are best explained by the process of self-scattering,  whether this process could also have an impact in the case of Bok globules remains to be investigated. 

To answer this question is important because only the polarization by dichroic emission or absorption of magnetically aligned dust grains allows us to draw conclusions about the underlying magnetic field structure, while polarization by scattering primarily provides information about  the dust grain size and distribution \citep[e.g.,][]{Brunngraeber2019}.  
In this context, it is important to note that the polarization pattern alone is not a unique indicator of the underlying polarization mechanism because polarization by scattering can generate patterns  that mimic the polarization patterns produced by a toroidal or poloidal magnetic field \citep{Yang2016, Brunngraeber2019}. \\
While previous numerical studies of self-scattering were mostly focussed on circumstellar disks \citep[e.g.,][]{Yang2017, Brunngraeber2020}, its role in the case of Bok globules has barely been addressed. However, a complete analysis of self-scattering in Bok globules is beyond the scope of this study. Instead, we limit ourselves to examining whether self-scattering plays a role in B335 at a wavelength of 214\,$\mu$m. 
For our analysis, we simulate self-scattering by spherical dust grains based on the temperature profile of the dust and consider the best-fit model parameter values for the dust and structure of B335 found in Section \ref{Section:Model_of_B335}. 

\begin{figure}[h]
   \centering
   \includegraphics[width=\hsize]{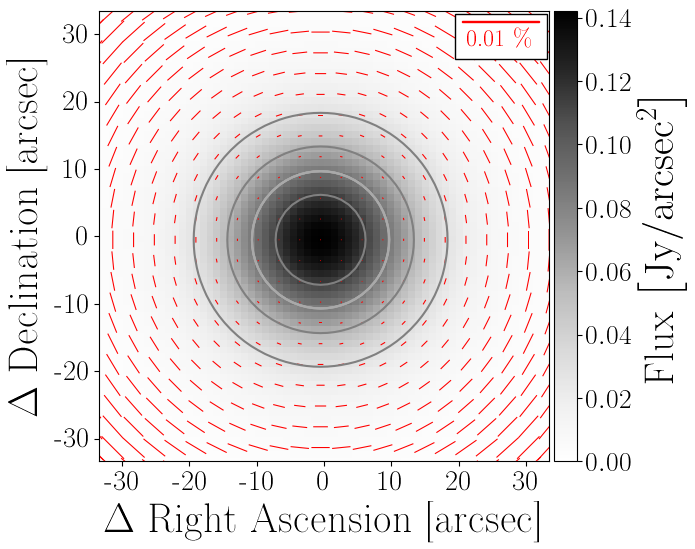}
      \caption{Synthetic polarization map of B335 at a wavelength of 214\,$\mu$m derived under the assumption of self-scattering. Total intensity is shown (re-emission and scattered light) with overlaid polarization vectors in red. Polarization due to self-scattering causes a centrosymmetric polarization pattern for our model. The resulting degrees of polarization for our self-scattering model are very low (see reference vector in the upper right corner of the polarization map). The isocontour lines mark 20\%, 40\%, 60\%, and 80$\%$ of the maximum intensity. The image has been convolved with a Gaussian profile with the SOFIA/HAWC+ band E FWHM of 18.2$''$.
              }
         \label{Figure:Intensity_Map_POLARIS_Self_Scattering}
\end{figure}

The resulting polarization pattern due to self-scattering shows a centrosymmetric pattern (see Fig. \ref{Figure:Intensity_Map_POLARIS_Self_Scattering}) and is qualitatively different from the observed pattern in Fig. \ref{Pol_Map_Only_Valid_Vectors}. Furthermore, the low optical depth ($\tau_{\text{214}\mu\text{m}} \approx 0.1$, see Sect. \ref{Section:Model_of_B335}) and the small grain size (compared to the observing wavelength) result in both low scattering efficiency and a low degree of scattered light polarization.
Consequently, the polarization degree is several magnitudes smaller than the observed one \citep{Kataoka2015}. Therefore, we can exclude a significant contribution of the effect of self-scattering as an explanation for the observed polarization pattern and polarization degree, and conclude that the measured polarization is caused by the emission of nonspherical magnetically aligned dust grains.

\subsection{Magnetic field structure and strength} \label{Section:Magnetic_field}
\begin{figure}
   \centering
   \includegraphics[width=\hsize]{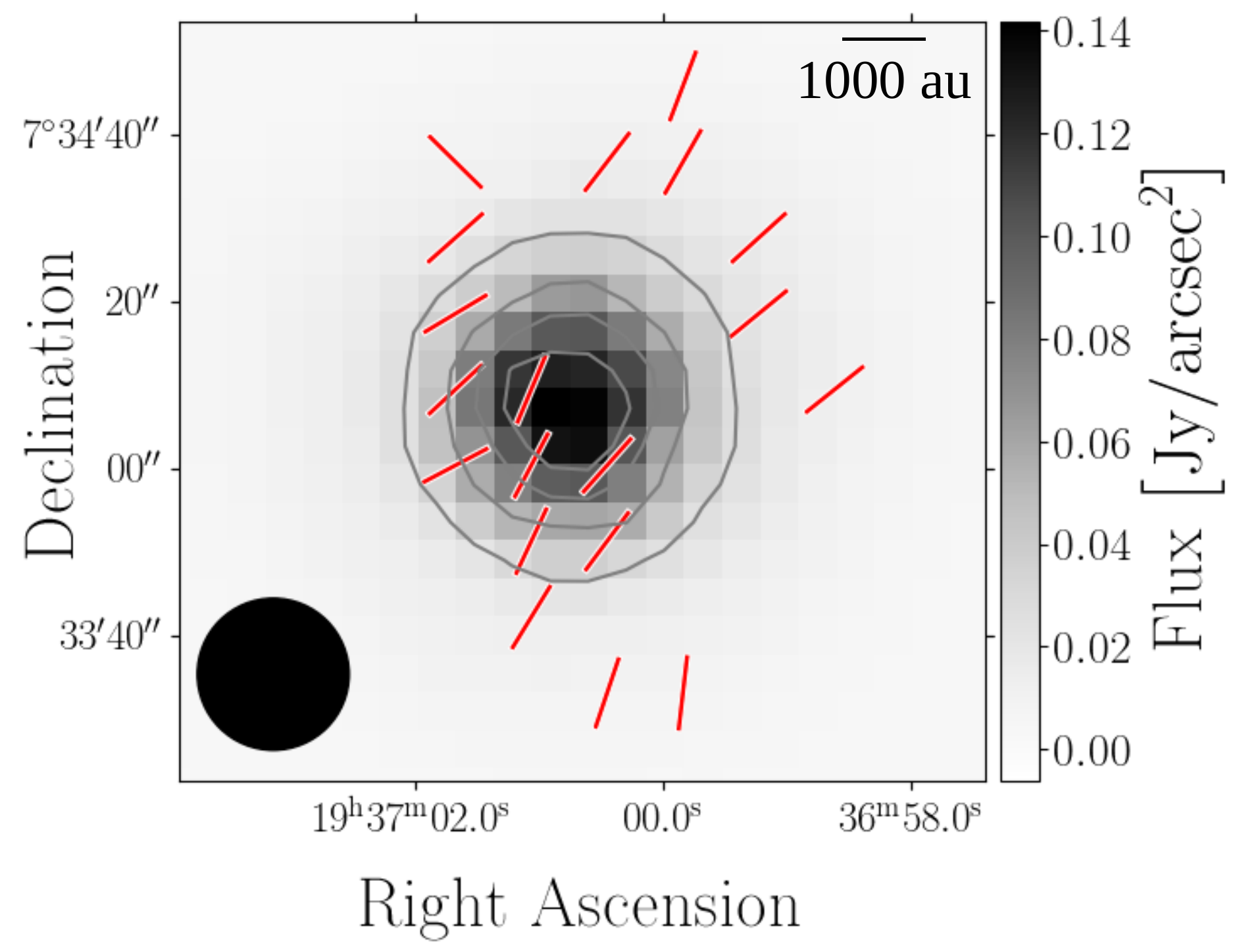}
      \caption{Intensity map of B335 with polarization vectors rotated by  90$^\circ$ to indicate the projected local magnetic field direction. The length of the vectors is arbitrary and constant and does not indicate the strength of the magnetic field. The beam size of 18.2$''$ is indicated in the lower left.
              }
         \label{Pol_Map_Only_Valid_Vectors_B_Field_Direction}
\end{figure}

As discussed in Sect. \ref{Section:Self_Scattering}, the measured polarization is caused by the emission of nonspherical magnetically aligned dust grains. In this case, the polarization vectors rotated by 90$^\circ$ indicate the projected magnetic field direction of B335 (Fig. \ref{Pol_Map_Only_Valid_Vectors_B_Field_Direction}).  As the polarization vectors show a uniform pattern, the resulting magnetic field structure is also uniform on the scales resolved by our observation\footnote{SOFIA/HAWC+ band E beam size is 18.2$''$, corresponding to 1820\,au}. To get an estimate for the magnetic field strength of B335, we apply the Davis-Chandrasekhar-Fermi method \citep{Davis1951, ChandrasekharFermi1953}:
\begin{equation} \label{Equation:DCF-method}
\hspace{3cm}  B = \sqrt{ \frac{4\pi}{3}  \rho_\text{gas}  } \: \frac{   v_\text{turb} }   {   \sigma_{\bar{\theta}}   }. \qquad \mathrm{[CGS]}
\end{equation}
Here, $ \rho_\text{gas}$ describes the gas density, $v_\text{turb}$ the turbulence velocity, and $\sigma_{\bar{\theta}}$ the standard deviation of the mean polarization angle $\bar{\theta}$ of the polarization vectors. One should keep in mind that, in the best case, the magnetic field should be estimated as a function of distance to the center because the density increases towards the core. However, due to an insufficient number of polarization vectors, we estimate the magnetic field strength based on the mean value, considering the gas density of our best-fit model (averaged over the 3\,$\sigma$ region) \mbox{$\rho_\text{gas}$ = 5.41 $\cdot$ 10$^{-18}$ g cm$^{-3}$.} The value for \mbox{$\sigma_{\bar{\theta}}$ =  26.8$^{+12.9}_{-6.5}$\,deg} is calculated using the 95$\%$ confidence interval for the standard deviation. We apply a turbulence velocity of 0.14\,cm s$^{-1}$ \citep{Frerking1987} and derive a magnetic field strength of 142$\pm 46\,\mu$G at 214$\,\mu$m.
\noindent The calculated magnetic field strength of B335 is in the same range as derived for other globules \citep[$\sim$100-300$\,\mu$G, see e.g.,][]{Wolf2003}. For  B335, \citet{Wolf2003} report a magnetic field strength of 134$^{+46}_{-39}$\,$\mu$G at 850$\,\mu$m.  As the polarization pattern at 214$\,\mu$m is more uniform than at 850$\,\mu$m \citep[see Fig. 1 in][]{Wolf2003}, the magnetic field strength, which was calculated on the basis of the far-IR data, is slightly higher.

\noindent Based on the derived magnetic field strength, it is possible to calculate the mass-to-flux ratio $M/\Phi$. This ratio indicates whether the core is in collapse or equilibrium. Following \cite{Crutcher2004}, we calculate the observed mass-to-flux ratio to the critical mass-to-flux ratio:
\begin{equation}
\frac{  (M/\Phi)_\text{observed} }{ (M/\Phi)_\text{crit} } = 7.6 \cdot 10^{-21} \: \frac{N(H_2)}{B} \: \cdot \: \frac{\mu\text{G}}{\text{cm}^2} .
\end{equation} 
Here, $N(H_2)$ describes the column density and $B$ the magnetic field strength. We use the value for $N(H_2)$  corresponding to our best-fit model (see Sect. \ref{Section:Model_of_B335}). We consider the average value of the column density inside a circle around the peak value with a radius of 3$\sigma$ and obtain \mbox{ $N(H_2)$ $\approx$ 6.28 $\cdot$ 10$^{22}$ cm$^{-2}$. } This yields a mass-to-flux ratio of $\sim$ 3.4, which indicates that the magnetic field is too weak to support against gravity by magnetic pressure alone; in other words, the core is slightly supercritical.

\subsection{Influence of the optical depth $\tau$ on the polarization pattern} \label{Section:Influence_optical_depth}
Various hypotheses exist that explain the decrease in polarization degree towards the center of Bok globules. One is based on the possible impact of dichroic absorption in addition to dichroic emission \citep{BrauerWolfReissl}. For this reason, and as this theory is relatively easy to verify, we analyze the influence of the optical depth on the polarization pattern in the following, based on our best-fit model derived in Sect. \ref{Section:Model_of_B335}.

\begin{figure}[h]
   \centering
   \includegraphics[width=\hsize]{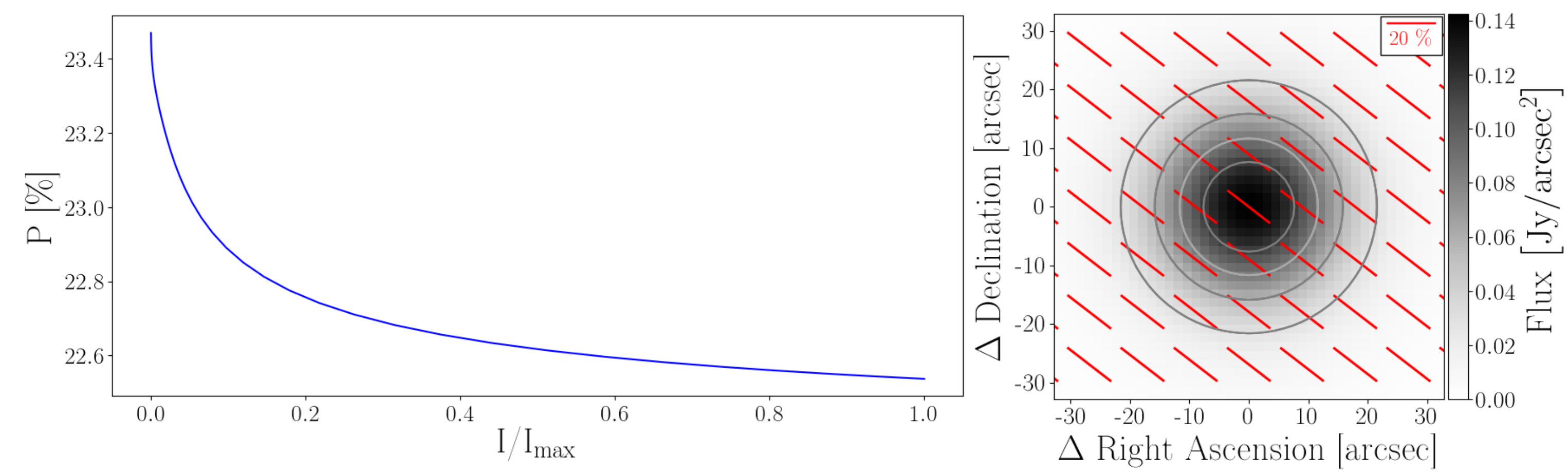}
      \caption{Correlation between polarization degree and normalized intensity (left) and synthetic intensity map (right) with overlaid polarization vectors for a Bok globule with \mbox{$M$ = 4.6\,M$_\odot$}  at $\lambda$ = 214\,$\mu$m. Perfectly aligned dust grains are assumed. The isocontour lines mark 20\%, 40\%, 60\%\ and 80$\%$ of the maximum intensity. The image has been convolved with a Gaussian with the SOFIA/HAWC+ band E FWHM of 18.2$''$.
              }
         \label{Scatter_Plot_Intensity_Map_Perfect_Alignment}
\end{figure}

In order to generate polarization maps based on our model, we apply a constant and homogenous magnetic field  for simplicity, assuming perfectly aligned dust grains. The direction of the magnetic field is based on our SOFIA/HAWC+ observation (see Fig. \ref{Pol_Map_Only_Valid_Vectors_B_Field_Direction}). We consider the same dust mixture as in Section \ref{Section:Model_of_B335}, but with oblate grains with an axis ratio $a/b$ = 2. Here, the radius of an oblate dust grain is given by the radius of an equal-volume sphere. The optical properties were calculated with a code by N. Voshchinnikov\footnote{http://www.astro.spbu.ru/DOP/6-SOFT/SPHEROID/1-SPH$\_$new/} based on \citet{Voshchinnikov1993}. See Fig. \ref{Scatter_Plot_Intensity_Map_Perfect_Alignment} (right) for the resulting (beam-convolved) polarization map based on our best-fit model.

As the derived optical depth for our model is only on the order of $\tau_{\text{214}\mu\text{m}} \approx 0.1$ towards the center, the dichroic extinction has little impact on the net polarization of B335. More specifically, the polarization degree decreases from the outer regions to the core by only a factor of $\sim$ 0.95 (see Fig. \ref{Scatter_Plot_Intensity_Map_Perfect_Alignment} left). The optical depth effect can therefore be excluded to explain the decrease of the polarization degree towards the dense regions of B335. As this effect cannot be excluded in general for other Bok globules,  whether or not the optical depth could have an impact on the formation of polarization holes or  can be excluded as a possible origin needs to be investigated, especially as this effect
has already been observed in the case of the young stellar object NGC1333 IRAS4A \citep{Ko2020}. The degree of polarization that is seen in this simple model is too high. This aspect is considered in the following section.

\subsection{Influence of the stellar source and the interstellar radiation field} \label{Influence_of_the_stellar_source_and_the_ISRF}
The derived polarization degree based on our simple model is significantly higher than observed. Therefore, the model is expanded below with regard to the alignment of dust grains.
A significant contribution of self-scattering to the observed polarization in B335 at 214\,$\mu$m can be excluded  (see Sect. \ref{Section:Self_Scattering}). The remaining mechanism is dichroic emission of magnetically aligned nonspherical grains. Multiple dust-alignment theories exist, and the most promising one relies on radiative torques \citep[RATs, e.g., ][]{Lazarian2007, Hoang2009}, which are considered in the following. Here, nonspherical dust grains align in the presence of a radiation field, i.e., a central star. For an overview of how the radiative torque theory is implemented in POLARIS, see for example \citet{Reissl2016, Reissl2020}. \\
For the analysis of grain alignment in the case of B335, both the central star and the omnipresent interstellar radiation field (ISRF) are considered.  
 We use the description of the ISRF provided by \citet{Mathis1983}, applying a scaling factor of 1.47 \citep{Seifried2020}. 
In the RATs framework, the degree of polarization is influenced by multiple parameters. The RAT efficiency $Q_\Gamma$, defined as
\begin{equation}
\hspace{2cm}  Q_\Gamma =  \begin{cases}

\quad Q_\Gamma^\text{ref}  & \quad \text{if} \quad \lambda \le 1.8\:a  \\ 

\quad  Q_\Gamma^\text{ref} \cdot \left( \frac{\lambda}{1.8\,a} \right)^{\alpha_Q}      &  \quad \text{otherwise,}

\end{cases}
\end{equation}
is particularly important \citep{Draine1996, Hoang2014}. The required parameters $Q_\Gamma^\text{ref}$ and $\alpha_Q$ depend on grain shape and material, but are not well constrained. The parameter $\alpha_Q$ is in the range of -2.6 to -4 \citep[][]{LazarianHoang2007, Herranen2019, Reissl2020}. We set $\alpha_Q$ to an average value of -3, while we adjust the parameter $Q_\Gamma^\text{ref}$  to fit the observations. The values for $Q_\Gamma^\text{ref}$ are assumed to vary between 0.01 and 0.4 \citep{LazarianHoang2007, Reissl2020}. 

\begin{figure*}
   \centering
   \includegraphics[width=1.0\textwidth]{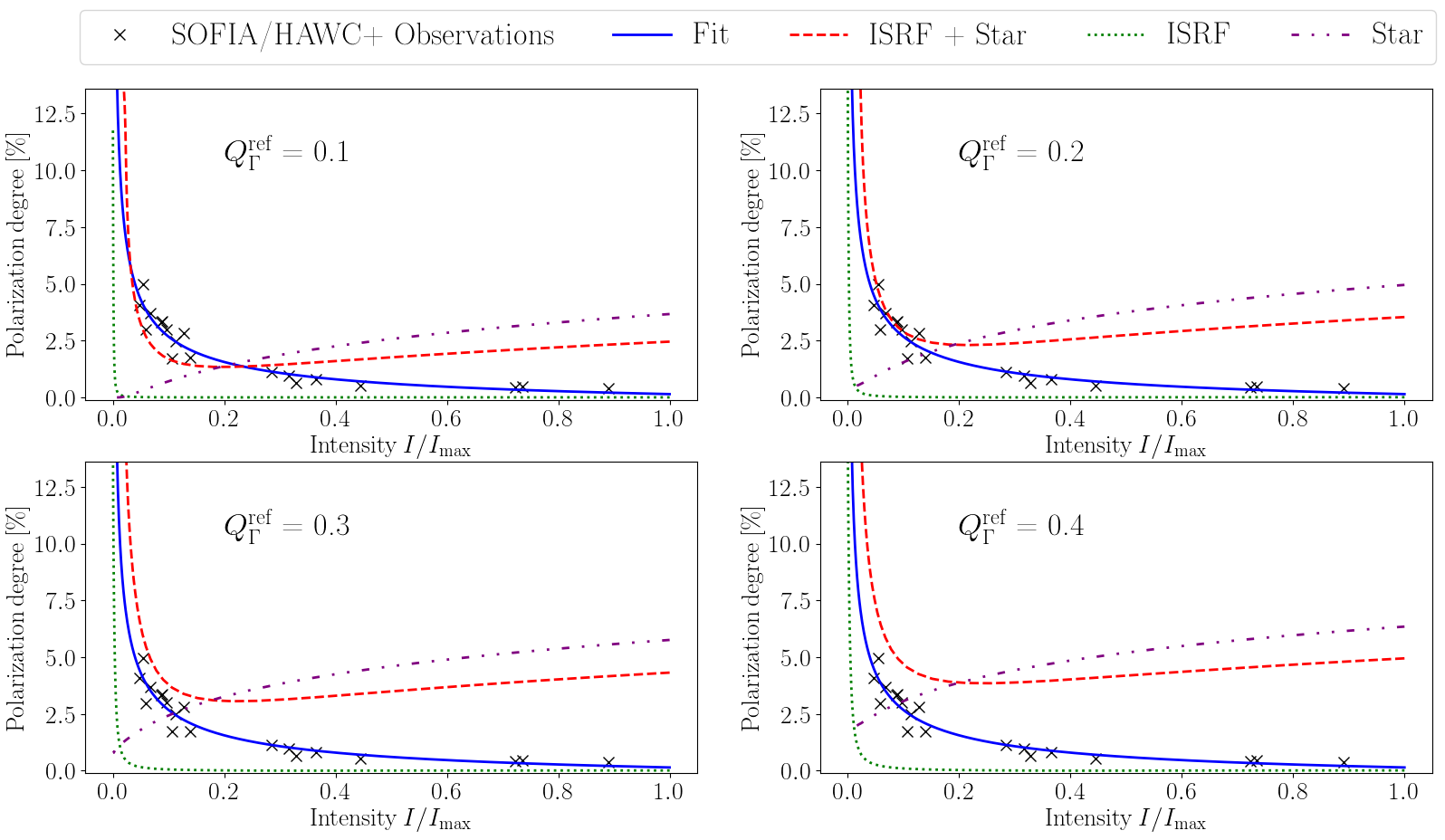}
      \caption{Correlation between polarization degree and normalized intensity for  three different cases of radiation sources and the SOFIA/HAWC+ observation of B335 at 214\,$\mu$m. Green: Radiation field based only on the ISRF; Purple: Radiation field based only on the star; Red: Radiation field based on the star and ISRF; Blue: Best-fit of our observation (see Fig. \ref{Scatter_Plot_B335}).
              }
         \label{Profile_Pol_Degree_Star_ISRF_Star_and_ISRF_All_Q_eff}
\end{figure*}

In the following, we consider the radiation fields of subsequent sources separately:
\begin{itemize}
\item a) Central star only, 
\item b) ISRF only and,
\item c) central star and ISRF combined.
\end{itemize}
We apply our best-fit model derived in Sect. \ref{Section:Model_of_B335}. The temperature map of B335 is calculated self-consistently on the basis of the aforementioned radiation source(s) and the dust distribution. Subsequently, on basis of the temperature maps, the re-emission maps are computed. In the final step, equivolume oblate dust grains with an axis ratio of 2:1 are considered.
Figure \ref{Profile_Pol_Degree_Star_ISRF_Star_and_ISRF_All_Q_eff} shows the correlation between polarization degree and normalized intensity for four different values of $Q_\Gamma^\text{ref}$ $\in$ (0.1, 0.2, 0.3, 0.4).

We find that the ISRF ensures alignment of the dust grains in the outer regions of the globule, but not in the inner region because the radiation does not penetrate deep enough. This is due to an overly high optical depth for the stellar contribution to the ISRF ($\lambda \lesssim 5\,\mu$m, $\tau_{1\mu \mathrm{m}, \mathrm{core}} \sim 120$), which is important for the alignment of the grains. However, the ISRF alone cannot explain the drop in the degree of polarization because it produces a significantly steeper curve. The high anisotropy and strength of the stellar radiation in the central region causes an increase in polarization degree towards the center, contradicting the observations of polarization holes. The combination of ISRF and star leads to a similar slope at the outer region to that seen in the observation. We find that a RAT effiency of $Q_\Gamma^\text{ref}$ $\approx$ 0.1 - 0.2 leads to a result that matches the observation at the outer region of the globule. These values are very close to the one determined by \citet{Reissl2020} in a study of grain alignment in the diffuse ISM, i.e., $Q_\Gamma^\text{ref}$ = 0.14. However, the decrease in the degree of polarization towards the center cannot be explained on the basis of our model.


\section{Discussion} \label{Section_Discussion}

\subsection{SOFIA/HAWC+ observation of B335 in the context of further polarimetric observations} \label{Section:HAWC_plus_observations_of_B335_in_the_context_of_further_polarimetric_observations}
\begin{figure*}
   \centering
   \includegraphics[width=1.0\textwidth]{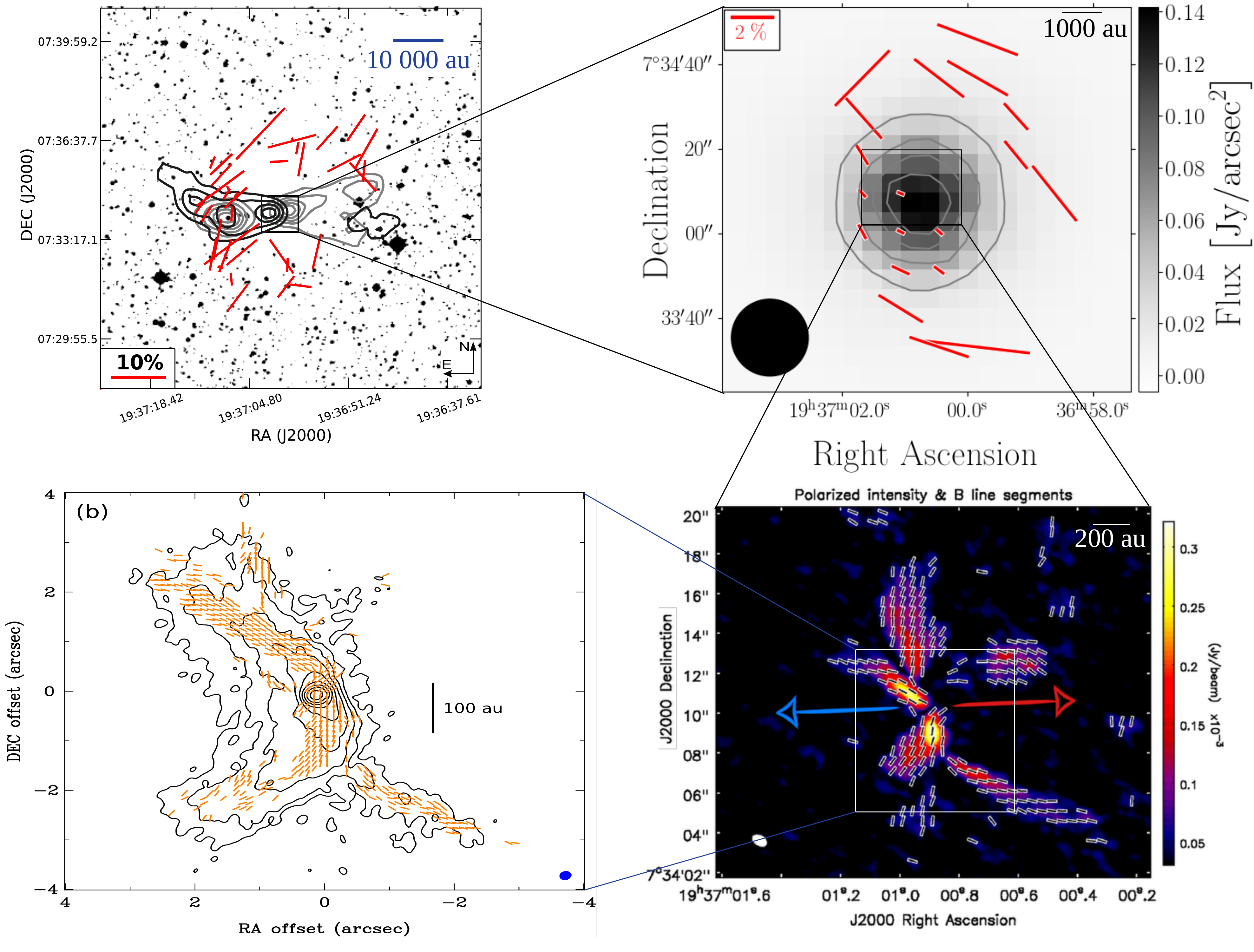}
      \caption{Polarization maps and corresponding magnetic field structure of B335. \textbf{Top-left:} Polarization map of B335 in the near-IR  \citep[Js band, 1.24\,$\mu$m, overlaid on a DSS intensity map;][]{Bertrang2014}. \textbf{Top-right:} Polarization map of B335 at 214\,$\mu$m (overlaid on the SOFIA/HAWC+ intensity map; this publication). \textbf{Bottom-right:}   Polarization map of B335 at 1300\,$\mu$m \citep[overlaid on the ALMA polarized intensity map; reuse of Figure 1 in][]{Maury2018}. Shown is the derived magnetic field direction. The arrows indicate the outflow direction. \textbf{Bottom-left:} Polarization map of B335 at 870\,$\mu$m \citep[overlaid on the ALMA contour map;][\copyright AAS. Reproduced with permission]{Yen2020}. Shown is the derived magnetic field direction.  }
         \label{Figure:Composition_Bertrang_HAWC_Maury}
\end{figure*}

B335 is one of the most observed star-forming regions and has been particularly well covered by polarimetric observations, ranging from the near-IR to submm/mm wavelength; see Fig. \ref{Figure:Composition_Bertrang_HAWC_Maury}.  Nevertheless, polarimetric results in  the far-IR are still missing. We give a short overview of selected observations in connection with the observation that we present here.  

\subsection*{Near-infrared, 1.24\,$\mu$m (Js Band):}

\citet{Bertrang2014} (see Fig. \ref{Figure:Composition_Bertrang_HAWC_Maury} top-left) report a largely well-ordered polarization pattern at the near-IR ($\bar{\theta}$ = 103.94$^\circ$ $\pm$ 5.01$^\circ$) at large scales ($\approx$ 10$^4$ - 10$^5$\,au). 
For the comparison of the magnetic field direction, we infer for the near-IR results that the polarization vectors are parallel to the magnetic field (dichroic absorption). Thus, the magnetic field direction is in a comparable range to our findings ($\bar{\theta}_B$ $\approx$ 140$^\circ$, Fig. \ref{Pol_Map_Only_Valid_Vectors_B_Field_Direction}). The magnetic field strength in the near-IR \citep[12-40$\,\mu$G,][]{Bertrang2014} is smaller by a factor of 5-15 than in the far-IR ($\sim$ 142$\,\mu$G). As the gas density is significantly lower on larger scales where the magnetic field is traced through observations at near-IR wavelengths, this finding can be attributed to the dependence of the magnetic field strength on the gas density in the flux-freezing limit \citep{Heitsch2005}.

\subsection*{870\,$\mu$m and 1.3\,mm:}
\citet{Maury2018} report that  the magnetic field structure at 1.3\,mm shows an ordered topology within the inner 
envelope, with a transition from a large-scale poloidal magnetic field in the outflow direction to a strongly pinched one in the equatorial direction (Fig. \ref{Figure:Composition_Bertrang_HAWC_Maury} bottom-right). Consistent with the 1.3\,mm observations, \citet{Yen2020} find that the magnetic field at 870\,$\mu$m (Fig. \ref{Figure:Composition_Bertrang_HAWC_Maury} bottom-left) changes from ordered to more complex and asymmetric structures within the inner 100 au of B335.  The variation of the polarization angles is significantly higher at 870\,$\mu$m and 1.3\,mm than at 214\,$\mu$m.  However, one has to keep in mind that a possibly more complex structure at 214$\,\mu$m is not resolved due to the lower resolution of SOFIA/HAWC+. The observed polarization degrees at wavelengths of 870\,$\mu$m and 1.3\,mm are in the range of $\sim$\,1-11$\%$, while for the 214\,$\mu$m (HAWC+) case the degree of polarization varies between $\sim$\,0.5\ and 5\,$\%$.  The derived magnetic field strength at 1.3\,mm is higher (300-3000 $\mu$G) than our result. In analogy to the possible role of flux-freezing discussed above, this finding may be explained by a strong inward drag of the magnetic field lines \citep{Maury2018}.

\subsection{Polarization hole in B335 at 214\,$\mu$m} \label{Section:Polarization_hole_in_B335}
The SOFIA/HAWC + band E (214\,$\mu$m) observation shows a decrease in the degree of polarization towards the center of B335 ("polarization hole", see Fig. \ref{Scatter_Plot_B335}). 
Our model can explain the behavior of the degree of polarization (P vs. I) in the outer regions of B335, but not in the core region. 
In the following we evaluate the various possible reasons for the occurrence of this polarization hole:

\begin{itemize}
\item[i)] Our model is based on RATs and a homogeneous magnetic field. While the assumption of a simple homogeneous magnetic field on large scales based on the SOFIA/HAWC+ observations is motivated by the rather homogeneous polarization pattern on that scale, it is not applicable on smaller scales ($\lesssim$ 1000\,au). High-resolution ALMA observations of B335 show that the magnetic field changes from a homogeneous structure to a more complex asymmetrical structure within the inner $\sim$ 100\,au \citep[see Fig. \ref{Figure:Composition_Bertrang_HAWC_Maury} bottom-left and bottom-right,][]{Maury2018,Yen2020}. SOFIA/HAWC+ cannot resolve these complex substructures. Beam-averaging over these complex magnetic field regions would result in a decrease in polarization degree \citep[e.g.,][]{Glenn1999, Matthews2005}.

\item[ii)] If the optical depth was higher than actually determined in our modeling approach (see Sect \ref{Section:Influence_optical_depth}), the increased impact of dichroic extinction would also explain the decrease in polarization degree \citep{BrauerWolfReissl}

\item[iii)] Our model consists of a mixture of silicate and graphite dust grains homogeneously distributed over the Bok globule. For our simulations, we assume that graphite grains do not align with the magnetic field \citep[e.g.,][]{Hildebrand1999, Hoang2015}. However, whether this assumption is justified is a matter of open debate \citep[e.g., ][]{Chiar2006, Lazarian2020}. Therefore, an increase in the abundance of graphite towards the center, as assumed by \citet{Olofsson2011}, would decrease the degree of polarization. 
However, in order to achieve a degree of polarization of $\approx$\,0.5$\%$ in the center of B335 for our model, we would need an unrealistically high abundance of graphite (> 80$\%$).

\item[iv)]  In addition, the decrease in the degree of polarization towards the center could also be explained with less aligned dust grains at the core \citep{Goodman1992, Creese1995}. With an alignment efficiency of $Q_\Gamma^\text{ref}$  = 0.01 \citep[smallest reasonable value according to the literature; ][see Sect. \ref{Influence_of_the_stellar_source_and_the_ISRF} for details]{LazarianHoang2007, Reissl2020}, the polarization degree at the core of B335 amounts to $\approx$ 0.5$\%$.  The polarization hole in B335 could therefore be explained by a decrease in the  alignment efficiency by an order of magnitude towards the center.

\item[v)]The polarization degree of thermal radiation depends on the absorption cross-sections of the long ($\perp$) and short ($\parallel$) axis of the dust grains \citep[$C_{\mathrm{abs}, \perp} - C_{\mathrm{abs}, \parallel}$, e.g.,][]{Li2008, Draine2009}. Therefore, less elongated dust grains with an axis ratio a/b $\ll$ 2 at the core of the Bok globule would result in a decrease in the degree of polarization.

\item[vi)] Recently, a new, potentially promising mechanism was proposed: radiative torque disruption \citep{Hoang2019b, Hoang2019a}. Here, based on the centrifugal force, spinning large dust grains are disrupted into smaller fragments. It was shown that this effect may result in a decrease in the polarization degree with increasing incident radiation field, explaining the occurrence of polarization holes towards protostars \citep{Hoang2020}. 

\end{itemize} 

The fact that high-resolution observations have shown a more complex magnetic field substructure suggests that this is the major contributor of the depolarization at the inner regions of B335.



\section{Conclusions} \label{Section_Conclusion}
Here we report our analysis of polarimetric observations of the Bok globule B335 at 214\,$\mu$m using SOFIA/HAWC+. 

\begin{itemize}
\itemsep 10pt

\item[1.] For the first time, we show that polarization holes in Bok globules can also occur in the far-IR range. The polarization degree decreases from $\sim$ 5$\%$ in the outer regions (distance of $\sim$ 50$''$, i.e., 5000\,au from the core
) to $\sim$ 0.5$\%$ at the core (angular resolution of the polarization map: 18.2$''$). 

\item[2.] The polarization vectors show a very uniform pattern across the Bok globule, with a mean polarization angle of 48$^\circ\pm$\,26$^\circ$. We determine the magnetic field strength to 142\,$\pm\,46\,\mu$G, which is in agreement with other magnetic field strength calculations of Bok globules. 

\item[3.] We determine the mass of B335 to be $\sim$ 4.6$\,$M$_\odot$ and a maximum dust grain size of $a_\text{max}$ = 2\,$\mu$m.

 The corresponding optical depth is not  high enough to explain the observed polarization hole with dichroic extinction ($\tau_{\text{214}\mu\text{m}} \approx 0.1$ towards the center). 

\item[4.] Our findings of the polarization pattern and the inferred magnetic field directions at large scales are in good agreement with the near-IR results (deviation is $\sim$ 40$^\circ$). The polarization degrees at submm/mm and near-IR wavelengths ($\sim$\,1-10\,$\%$) are higher than in the far-IR ($\sim$\,0.5-5\,$\%$). From large scales (10$^3$ - 10$^5$au) to small scales ($\sim$10$^2$\,au), the magnetic field structure changes from a mostly uniform to a more complex structure.

\item[5.] A combination of the interstellar radiation field and the central star as radiation sources is able to explain the decrease in polarization degree at the outer regions ($\approx$\,10$^4$\,au from the core) of B335. 
However, the model fails to explain the low polarization degree within the inner 5\,000\,au. A complex magnetic field structure, as shown by ALMA \citep{Maury2018, Yen2020} but unresolved with SOFIA/HAWC+, provides a possible explanation for the decrease in polarization degree at the inner region.
\end{itemize}

\section*{Acknowledgements}
We thank Robert Brauer and Stefan Heese for data aquisition and helpful discussion. This paper is based on observations made with the NASA/DLR Stratospheric Observatory for Infrared Astronomy (SOFIA). SOFIA is jointly operated by the Universities Space Research Association, Inc. (USRA), under NASA contract NNA17BF53C, and the Deutsches SOFIA Institut (DSI) under DLR contract 50 OK 0901 to the University of Stuttgart.  Acknowledgement: N.Z. and S.W. acknowledge the support by 50OR1910. R. B. gratefully receives support through the DFG grant WO857/18-1. N.Z. thanks F. Santos for helpful comments.

\bibliographystyle{aa} 
\bibliography{lit}

\onecolumn
\begin{appendix}
\section{Polarization measurements of B335 using SOFIA/HAWC+ band E}

\begin{table*} [h]
\centering
\caption{SOFIA/HAWC+ band E polarization measurements for the Bok globule B335.  $P$ represents the polarization degree and $\theta$ the polarization angle. $\sigma_P$ and $\sigma_\theta$ denote the corresponding uncertainties. 
The values for $\theta$ and $\sigma_\theta$  describe the polarization measurements and not the derived magnetic field direction. }\label{Table:SOFIA_Detections}
\centering
\begin{tabular}{cccccc}
\hline\hline
\rule{0pt}{3ex}
Dec & RA & $P$ & $\sigma_P$ & $\theta$ & $\sigma_\theta$ \\
 & & (\%) & (\%) & ($\degr$) & (\degr)  \\  \hline \hline \\
7.0$^\circ$ 33.0$'$ 33.08$''$ &       19h 36 min 59.83s &  5.0 & 1.5  &   83.4  &   8.7 \\
7.0$^\circ$ 33.0$'$ 33.08$''$ &       19h 37 min 00.45s  & 3.0 & 1.1   &  71.4   &  10.4 \\ 
7.0$^\circ$ 33.0$'$ 42.18$''$ &       19h 37 min 01.06s & 2.5 & 0.9   &  58.5  &   9.9 \\
7.0$^\circ$ 33.0$'$ 51.28$''$ &       19h 37 min 01.06s & 1.0 & 0.3   &  64.9  &   9.4 \\
7.0$^\circ$ 33.0$'$ 51.28$''$ &       19h 37 min 00.45s & 0.6 & 0.2  &   53.3  &   10.6 \\
7.0$^\circ$ 34.0$'$ 00.38$''$ &       19h 37 min 00.45s & 0.5 & 0.2   &  48.1   &  11.1 \\ 
7.0$^\circ$ 34.0$'$ 00.38$''$ &       19h 37 min 01.06s & 0.4 & 0.1   &  62.3  &   8.3 \\
7.0$^\circ$ 34.0$'$ 00.38$''$ &       19h 37 min 01.67s & 0.8 & 0.3   &  27.8  &   10.5 \\
7.0$^\circ$ 34.0$'$ 09.48$''$ &       19h 37 min 01.67s & 0.5 & 0.2   &  42.9  &   10.4  \\
7.0$^\circ$ 34.0$'$ 09.48$''$ &       19h 37 min 01.06s & 0.4 & 0.1   &  67.2  &   8.2 \\
7.0$^\circ$ 34.0$'$ 18.58$''$ &       19h 37 min 01.67s & 1.1 & 0.4   &  30.9  &   10.3 \\
7.0$^\circ$ 34.0$'$ 27.68$''$ &       19h 37 min 01.67s & 2.8 & 0.8   &  41.6  &   8.3 \\
7.0$^\circ$ 34.0$'$ 36.78$''$ &       19h 37 min 01.67s & 3.7 & 1.3   &  -44.5  &   10.0 \\
7.0$^\circ$ 34.0$'$ 36.78$''$ &       19h 37 min 00.45s & 3.0 & 1.1   &  52.4  &   10.3 \\
7.0$^\circ$ 34.0$'$ 36.78$''$ &       19h 36 min 59.83s & 3.3 & 1.1  &   60.3  &   9.0 \\
7.0$^\circ$ 34.0$'$ 45.88$''$ &       19h 36 min 59.83s & 4.1 & 1.6  &   69.2  &   10.7 \\
7.0$^\circ$ 34.0$'$ 27.68$''$ &       19h 36 min 59.22s & 1.7 & 0.7  &   42.2  &   11.5 \\
7.0$^\circ$ 34.0$'$ 18.58$''$ &       19h 36 min 59.22s & 1.7 & 0.7  &   39.3  &   11.3 \\
7.0$^\circ$ 34.0$'$ 09.48$''$ &       19h 36 min 58.61s & 3.4 & 1.2   &  38.6  &   10.1 \\

\hline\hline
\end{tabular}
\end{table*}

\section{Overview: Observed and theoretical fluxes for B335 at different wavelengths}

\begin{table*} [h]
\begin{center}
\caption{Observed and simulated peak fluxes I$^{\text{p}}$ with POLARIS for maximum grain sizes \mbox{$a_\text{max} \in$ [0.25, 50]\,$\mu$m} and globule masses $M_\text{gas} \in$ [1, 5]\,M$_\odot$.  }\label{Table:Overview_Comparison_Observation_Simulation}
\centering
\begin{tabular}{cccccccc}
\hline\hline
 
\rule{0pt}{1ex}
{\small $\lambda$ } & {\small I$^{\text{p}}$ Obs. } & {\small I$^{\text{p}}$ (0.25$\,\mu$m) }& {\small I$^{\text{p}}$ (2$\,\mu$m) } & {\small I$^{\text{p}}$ (3$\,\mu$m) } & {\small I$^{\text{p}}$ (10$\,\mu$m) } & {\small I$^{\text{p}}$ (50$\,\mu$m) } \\
{\small ($\mu$m) } & {\small (Jy/beam) } & {\small (Jy/beam) } & {\small (Jy/beam) } & {\small (Jy/beam) } & {\small (Jy/beam) } & {\small (Jy/beam) } \\ 
&  & 5.0\,M$_\odot$ & 4.6\,M$_\odot$ & 4.4\,M$_\odot$ & 3.2\,M$_\odot$ & 1.0\,M$_\odot$   \\
 \hline \hline \\
1300 &  0.64 $\pm$ 0.07$^a$ &   0.25 &  0.42 &  0.59 &  0.95 &  0.40  \\
1100 &  0.68 $\pm$ 0.07$^a$  &  0.40 &  0.65 &  0.89 &  1.32 &  0.57 \\ 
850 &   1.1 $\pm$       0.33$^b$ &  0.72 &      1.13 &  1.46 &  1.89 &  0.83  \\
800 &   1.3 $\pm$       0.21$^a$ & 0.75 &       1.14 &  1.47 &  1.85 &  0.81  \\
600 &   4 $\pm$ 0.85$^a$ &  3.07 &      4.3 &   5.19 &  5.80 &  2.73  \\
450 &   4.2 $\pm$       0.84$^b$  & 2.52 &      3.3 &   3.85 &  3.85 &  1.95  \\ 
214 &   36.87 $\pm$ 0.02$^c$ &  37.36 & 36.72 & 37.02 & 37.38 & 36.90  \\
\hline  \rule{0pt}{3ex}
$\chi^2$ & & 1.61 & 0.35 & 0.59  & 2.40 & 1.97 \\  
\hline\hline 
\end{tabular} \\ 
\vspace{0.1cm} \textbf{Notes:} $^a$: \citet{Chandler1990}, beam sizes are 21$''$, 19$''$, 13$''$ and 19$''$ for 1300$\,\mu$m, 1100$\,\mu$m, 800$\,\mu$m and 600$\,\mu$m, respectively. \\ $^b$: \citet{Launhardt2010}, beam sizes are 8.6$''$ for 450$\,\mu$m and 14.9$''$ for 850$\,\mu$m. \\ $^c$: this paper, beam size is 18.2$''$.\\ The 2\,mm observation of \citet{Chandler1990} was excluded as no error limits were specified. 
\end{center}
\end{table*}

\end{appendix}

\end{document}